\documentclass[fleqn,twoside]{article}
\usepackage{espcrc2,epsfig}

\usepackage{graphicx}
\usepackage[figuresright]{rotating}

\newcommand{\AmS}{{\protect\the\textfont2
  A\kern-.1667em\lower.5ex\hbox{M}\kern-.125emS}}

\title{Towards absolute neutrino masses}

\author{Petr Vogel\address[MCSD]{Kellogg Radiation Laboratory 106-38, \\ 
        Caltech, Pasadena, CA 91125, USA}%
        \thanks{Support by the organizers of NOW2006 is gratefully acknowledged. 
The work was supported in part under U.S. DOE contract DE-FG02-05ER41361.}
       }
       
\begin{document}

\begin{abstract}
Various ways of determining the absolute neutrino masses are briefly
reviewed and their sensitivities compared. The apparent tension between the
announced but unconfirmed observation of the $0\nu\beta\beta$ decay and
the neutrino mass upper limit based on observational cosmology is used
as an example of what could happen eventually. The possibility of a
``nonstandard'' mechanism of the $0\nu\beta\beta$ decay is stressed and
the ways of deciding which of the possible mechanisms is actually 
operational are described. The importance of the $0\nu\beta\beta$ 
nuclear matrix elements is discussed and their uncertainty estimated.
\vspace{1pc}
\end{abstract}

\maketitle

\section{Generalities}

Thanks to the recent triumphs of neutrino physics we know that
neutrinos are massive fermions and that they are mixed, i.e., that
the neutrino flavor ($\nu_e, \nu_{\mu}, \nu_{\tau}$) 
is not a conserved quantity.
We also know, with a reasonable accuracy (some better than other),
the three mixing angles and the magnitudes of mass square differences
$\Delta m_{ij}^2 = m_i^2 - m_j^2$. 

These discoveries represent the first deviations from the Standard Model
of particle physics that postulated massless neutrinos and conservation
of the individual as well of the total lepton numbers. Thus, a new all
encompassing theory, sometimes called the ``New Standard Model'', should
be formulated. In order to delineate
a path to it, several additional questions ought to be answered. Among them, two
are the topic of this talk: ``Are neutrinos Majorana or Dirac fermions?''
and ``What is the absolute neutrino mass scale?''

The list below summarizes the methods currently used for neutrino mass
determination and their estimated sensitivities.
\begin{itemize}
\item {\it Neutrino oscillations:} Only mass squared differences, 
sometimes
only their absolute value, $\Delta m_{ij}^2 = m_i^2 - m_j^2$, are determined.
The two different $\Delta m^2$ values are  
$|\Delta m_{atm}^2| = (1.9 - 3.0) \times 10^{-3}$ eV$^2$ and
$\Delta m_{sol}^2 = 8.0_{-0.3}^{+0.4} \times 10^{-5}$ eV$^2$.
This range and indicated error bars show the present sensitivity.
The accuracy will undoubtedly improve soon, particularly for
$\Delta m_{atm}^2$. This mass determination is independent on the
charge conjugation properties of neutrinos.
\item {\it Ordinary beta decay:} The quantity determined or constrained
is $\langle m_{\beta} \rangle^2 = \Sigma_i m_i^2 |U_{ei}|^2$. Present limit
on $\langle m_{\beta} \rangle$ is $\sim$ 2 eV. The ultimate sensitivity
appears to be $\sim$ 0.2 eV. Again, independent on the Majorana or Dirac nature of
neutrinos.
\item {\it Observational cosmology:} The quantity determined or constrained
is $M = \Sigma_i m_i$. The sensitivity is at present model dependent,
but probably will eventually reach $\sim$ 0.1 eV. 
Again, independent on the Majorana or Dirac nature of
neutrinos.
\item {\it Double beta decay:} The quantity determined or constrained
is $\langle m_{\beta} \rangle = |\Sigma_i m_i |U_{ei}|^2 e^{i \alpha_i}|$,
where the Majorana phases $\alpha_i$ are at present totally unknown.
The sensitivity of the method in the near term is $\sim$ 0.1 eV, and in a longer 
term (next ten years or so) $\sim$ 0.01 eV. The $0\nu\beta\beta$ decay
exists only for Majorana neutrinos.
\end{itemize}

Note that other, sometimes conceptually simpler, methods of neutrino mass
determination cannot reach competitive sensitivities. For example, the
time-of-flight would use the time delay of massive neutrinos,
compared to massless particles, traveling a distance $D$, 
$\Delta t(E) = 0.514 (m/E)^2 D$ s, where $m$ is in eV, $E$ in MeV,
and $D$ in units of 10 kpc. For a galactic supernova various analyzes
suggest sensitivity $\sim$ 10-20 eV for this method.

The two body weak decays, like decay
of a pion at rest, $\pi^+ \to \mu^+ + \nu_{\mu}$, 
can be also used if the muon energy 
is determined. In that case 
$m_{\nu}^2 = m_{\pi}^2 + m_{\mu}^2 - 2 m_{\pi} E_{\mu}$. However, since 
this is a difference of two very large numbers, the present sensitivity
is only $\sim$ 170 keV with little hope for a substantial improvement.   

\begin{figure}[htb]
\centerline{\psfig{file=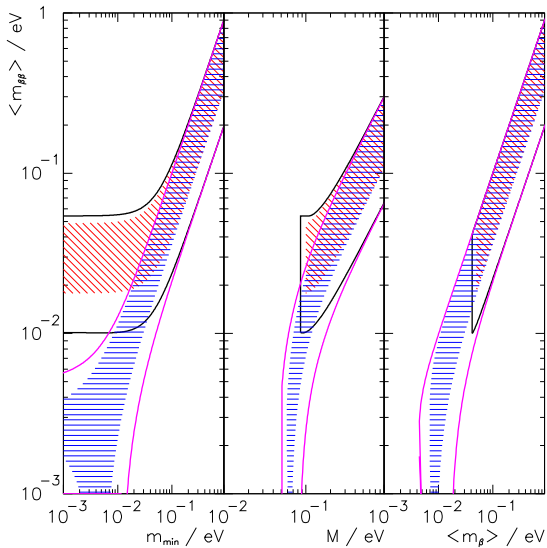,width=8.0cm}}
\caption{{\small Dependence of $\langle m_{\beta\beta} \rangle$
 on the mass of the lightest neutrino $m_{min}$, and on $M$ and
 $\langle m_{\beta} \rangle$.
 The irreducible
 width of the hatched areas is due to the unknown Majorana phases.
 The lines take into account the current uncertainties in the
 oscillation parameters; they will shrink as the accuracy improves. The two sets
 of curves
 correspond to the normal and inverted hierarchies.}}
\label{fig_fig1}
\end{figure}

The various neutrino mass dependent quantities are related, as shown in
Fig.\ref{fig_fig1}. Note that  a determination of $\langle m_{\beta\beta} \rangle$, 
even when combined
with the knowledge of $M$ and/or $\langle m_{\beta} \rangle$ does not allow,
in general, to distinguish between the normal and inverted mass orderings. This is
a consequence of the fact that the Majorana phases are unknown. In regions in
Fig. \ref{fig_fig1} where the two hatched bands overlap it is clear that two solutions
with the same  $\langle m_{\beta\beta} \rangle$ and the same $M$
(or the same $\langle m_{\beta} \rangle$) exist and cannot be distinguished.
On the other hand, obviously, if one can determine that
$\langle m_{\beta\beta} \rangle \ge$ 0.1 eV we would conclude that the
mass pattern is degenerate.  And in the so far hypothetical case
that one could show that $\langle m_{\beta\beta} \rangle \le$
0.01 - 0.02 eV, but nonvanishing
nevertheless, the normal hierarchy would be established.

\section{Current situation}

\begin{figure}[htb]
\centerline{\psfig{file=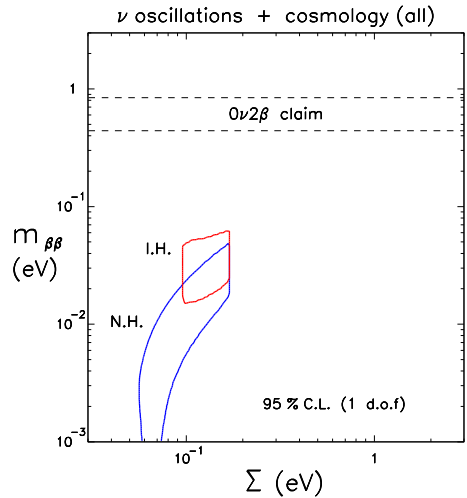,width=6.0cm}}
\caption{ {\small Apparent tension between the claim of the $0\nu\beta\beta$
decay discovery and the upper limit on the sum $\Sigma$
of the neutrino masses based on the 
observational cosmology (from \cite{fogli}). I.H. and N.H. mean inverted
and normal hierarchies.}} 
\label{fig_fig2}
\end{figure}

At present, some information exists on the degenerate mass region. I use this
as an avenue to discuss what can eventually happen, and what it might mean.
The results of the WMAP mission\cite{wmap} (3 years data), combined with
other observations\cite{Seljak} (Sloan Survey and in particular 
the Lyman-$\alpha$ forest analysis) restrict the sum of the neutrino masses
to about $\Sigma m_i \le$ 0.2 eV if all the data are combined. On the other
hand, a recently claimed (and as yet unconfirmed) discovery of the $0\nu\beta\beta$
decay\cite{Klapdor} would indicate that $\langle m_{\beta\beta} \rangle \ge$ 0.4 eV.
Putting these two indications together\cite{fogli} suggests an inconsistency
as shown in Fig.\ref{fig_fig2}.   

Leaving aside the all important question whether the $0\nu\beta\beta$ decay
experimental evidence will withstand further scrutiny and whether the cosmological
constraints are reliable and model independent, let us discuss possible scenarios
suggested by the comparison illustrated in Fig.\ref{fig_fig2}.

What can happen once all evidence becomes available:
\begin{enumerate}
\item Both neutrino mass determinations will yield a positive 
and consistent result, i.e., both results will intersect at the allowed
band and both will suggest the degenerate neutrino mass pattern. Such results
will be relatively readily accepted, even though many theorists do not
expect the degenerate scenario.
\item Future $0\nu\beta\beta$ decay experiments will not find a positive evidence
(i.e., the present claim will be shown to be incorrect), but the observational
cosmology or/and the study of tritium $\beta$ decay 
will find evidence for the degenerate mass pattern. This is the 
situation exactly opposite to the one depicted in Fig.\ref{fig_fig2}. 
This will
be also, albeit reluctantly, accepted and would indicate that neutrinos are Dirac,
not Majorana particles.
\item The depicted situation is confirmed. The positive evidence of $0\nu\beta\beta$
decay is confronted with a lack of confirmation from observational cosmology.
What then? Is there a possible scenario that would accommodate this situation?
\end{enumerate} 

\section{Mechanism of $0\nu\beta\beta$ decay}

The answer is yes and deserves a more detailed explanation. In fact, this can
happen for two reasons. Possibility 1): The $0\nu\beta\beta$ decay is 
not caused by the exchange 
of a light Majorana neutrino but by another mechanism. Hence the extraction
of $\langle m_{\beta\beta} \rangle$ from the lifetime is not possible.
Possibility 2):
Even though the $0\nu\beta\beta$ decay is caused by the exchange 
of a light Majorana neutrino the relation between the lifetime and
$\langle m_{\beta\beta} \rangle$ is different than used so far, since the 
nuclear matrix elements are highly uncertain.

In order to further discuss the point 1) above, note that
besides the exchange of a light Majorana neutrino $0\nu\beta\beta$ decay can be
caused by the exchange of various hypothetical heavy particles in 
particle physics models that
contain Lepton Number Violation (LNV). It turns out that the confusion
about the possible mechanism can occur if the scale of such heavy particles
is $\Lambda \sim$ 1 TeV. (Smaller scales are already excluded,
much larger ones lead to unobservably long lifetimes.) 
If the  $0\nu\beta\beta$ decay is observed, how can 
we tell which mechanism is responsible?

Generally, observation of the $0\nu\beta\beta$ decay, even of the single electron
spectrum and/or the angular distribution of the electrons, does not allow one to
determine the mechanism responsible for the decay. It has been suggested in Ref.\cite{lfv}
that the observation of the Lepton Flavor Violation (LFV) could be used as a ``diagnostic tool''
for that purpose.

The discussion is
concerned mainly with the branching ratios $B_{\mu \rightarrow e \gamma} = \Gamma
(\mu \rightarrow e \gamma)/ \Gamma_\mu^{(0)}$ and $B_{\mu \to e} =
\Gamma_{\rm conv}/\Gamma_{\rm capt} $, where $\mu \to e \gamma$ is
normalized to the standard muon decay rate $\Gamma_\mu^{(0)} = (G_F^2
m_\mu^5)/(192 \pi^3)$, while $\mu \to e$ conversion 
for the $\mu^-$ on an atomic orbit of a nucleus is normalized to
the corresponding capture rate $\Gamma_{\rm capt}$. The main diagnostic tool in the
analysis is the ratio
\begin{equation}
{\cal R} = B_{\mu \to e}/B_{\mu \rightarrow e \gamma} ~,
\end{equation}
and the relevance of our observation relies on the potential
for LFV discovery in the forthcoming experiments  MEG~\cite{MEG}
($\mu \to e \gamma$) and MECO~\cite{MECO} 
($\mu \to e$ conversion)\footnote{Even though MECO
experiment was recently canceled, proposals
for experiments with similar sensitivity exist elsewhere.}.

As explained in \cite{lfv} if the ratio ${\cal R}$ is $\sim 10^{-(2-3)}$
one expects that the  $0\nu\beta\beta$ decay is caused by the
exchange of a light Majorana neutrino and hence the decay rate is
proportional to $\langle m_{\beta\beta} \rangle^2$. On the other hand,
observation of  ${\cal R} \gg 10^{-2}$ could signal non-trivial LNV
dynamics at the TeV scale, whose effect on $0 \nu \beta \beta$ has to
be analyzed on a case by case basis. Therefore, in this scenario no
definite conclusion can be drawn based on LFV rates.
In addition, non-observation of LFV in muon processes in forthcoming
experiments would likely imply that the scale of non-trivial LFV and
LNV is  above a few TeV, and thus $\Gamma_{0 \nu \beta
\beta} \sim \langle m_{\beta \beta} \rangle^2$.

The conclusion above, with some important caveats, was reached in \cite{lfv}
by analyzing two phenomenologically viable models
that incorporate LNV and LFV at low scale, the left-right symmetric model
and the R-parity violating supersymmetry.
However, it is likely that the basic mechanism at work in
these illustrative cases is  generic: low scale LNV interactions
($\Delta L = \pm 1$ and/or $\Delta L= \pm 2$), which in general
contribute to $0 \nu \beta \beta$, also generate sizable contributions
to $\mu \to e$ conversion, thus enhancing this process over $\mu \to e
\gamma$.

\section{Nuclear matrix elements}

If indeed the exchange of a light Majorana neutrino is responsible for the $0\nu\beta\beta$ decay,
the half-life and the effective mass are related by
\begin{equation}
\frac{1}{T_{1/2}^{0\nu}} = G^{0\nu}(Q,Z) |M^{0\nu}|^2 \langle m_{\beta\beta} \rangle^2 ~,
\label{eq_rate}
\end{equation}
where $G^{0\nu}(Q,Z)$ is a phase space factor that depends
on the transition $Q$ value and through
the Coulomb effect on the emitted electrons on the nuclear charge $Z$ and that can be
easily and accurately calculated, and $M^{0\nu}$ is the nuclear matrix element that can be
evaluated in principle, although with a considerable uncertainty.

It follows from eq.(\ref{eq_rate}) that 
(i) values of the nuclear matrix elements $M^{0\nu}$ 
are needed in order to extract the effective neutrino mass from
the measured $0\nu\beta\beta$ decay rate, and (ii) any uncertainty in  $M^{0\nu}$
causes a corresponding and equally large uncertainty in the
extracted $\langle m_{\beta \beta} \rangle$ value. Thus, the issue of an accurate
evaluation of the nuclear matrix elements attracts considerable attention
and in its extreme form can explain the situation depicted in Fig.\ref{fig_fig2}.

Common to all methods of calculating $M^{0\nu}$ 
is the description of the nucleus as a system of nucleons bound
in the mean field and interacting by an effective residual interaction. The used methods
differ as to the number of nucleon orbits (or shells and subshells) included in
the calculations and
the complexity of the configurations of the nucleons in these orbits.
The two basic approaches used so far
for the evaluation of the nuclear matrix elements for both the $2\nu$ and $0\nu$
$\beta\beta$ decay modes
are the Quasiparticle Random Phase Approximation (QRPA) and the nuclear shell model (NSM).
They are in some sense complementary; QRPA uses a larger set of orbits, but truncates
heavily the included configurations, while NSM can include only a rather small set
of orbits but includes essentially all possible configurations. NSM also can be tested
in a considerable detail
by comparing to the nuclear spectroscopy data; in QRPA such comparisons are much
more limited.

Since the calculations using QRPA are much simpler, an overwhelming majority
of the published calculations uses that method. There are suggestions to use the
spread of these published values of   $M^{0\nu}$ as a measure of uncertainty\cite{BM04}.
Following this, one would conclude that the uncertainty is quite large, a factor of three
or as much as five. But that way of assigning the uncertainty is questionable. Using all
or most of the published values of $M^{0\nu}$ means that one includes calculations
of uneven quality. Some of them were devoted to the tests of various approximations,
and concluded that they are not applicable. Some insist that other data, like the
$M^{2\nu}$, are correctly reproduced, other do not pay any attention to such test.
Also, different forms of the transition operator $\hat{O}^{0\nu}$ are used, in particular
some works include approximately the effect of the short  range nucleon-nucleon repulsion,
while others neglect it.

\begin{figure}[htb]
\centerline{\psfig{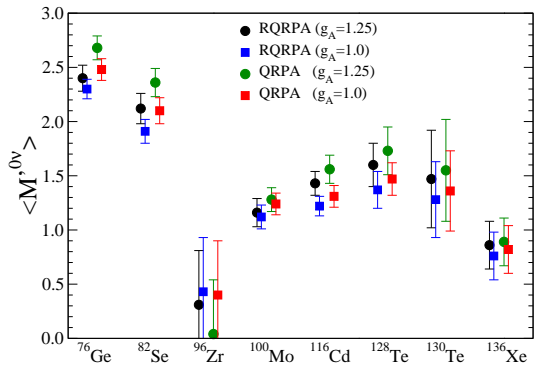}}
\caption{{\small Nuclear matrix elements and their variance for the indicated approximations
(see Ref.\cite{rodin}).}} 
\label{fig_fig3}
\end{figure}

In contrast, in Ref.\cite{rodin} an assessment of uncertainties
in the matrix elements $M^{0\nu}$ inherent in the QRPA was made, and it was concluded that
with a consistent treatment the uncertainties are much less, perhaps only about 30\%
(see Fig.\ref{fig_fig3}).
That calculation uses the known $2\nu$ matrix elements in order to adjust the
most important free parameter, the effective proton-neutron interaction constant.
There is a lively debate in the nuclear structure theory community,
beyond the scope of this talk, about this conclusion.

It is of interest also to compare the resulting matrix elements
of Rodin {et al.}\cite{rodin} based
on QRPA and its generalizations,
and those of the available most recent NSM evaluation\cite{Poves}.
Note that the operators used in NSM evaluation do not include the induced nucleon currents
that in QRPA reduce the matrix element by about 30\%. The QRPA\cite{rodin}
and NSM\cite{Poves} $M^{0\nu}$ are compared in Table \ref{tab_nme}. In the last column
the NSM matrix elements are reduced by 30\% to approximately account for the missing terms
in the operator, and to make the comparison more meaningful. With this reduction,
it seems that QRPA results are a bit larger in the lighter nuclei and a bit smaller in the heavier
ones than the NSM results, but basically within the 30\% uncertainty estimate.
Once the NSM calculations for the intermediate mass nuclei $^{96}$Zr, $^{100}$Mo
and $^{116}$Cd become available, one can make a more meaningful comparison of the two
methods.

\begin{table}
\begin{center}
\caption{{\small Comparison of the calculated nuclear matrix elements $M^{0\nu}$ using the QRPA
method\cite{rodin} and the NSM\cite{Poves}. In the last column the NSM values are reduced,
divided by 1.3, to account approximately for the effects of the induced nucleon currents.
} }
\vspace{0.2cm}
{\begin{tabular}{|c|c|c|c|}
\hline
Nucleus & QRPA & NSM & NSM/1.3 \\
\hline
$^{76}$Ge & 2.3-2.4 & 2.35 & 1.80 \\
$^{82}$Se & 1.9-2.1 & 2.26 & 1.74 \\
$^{96}$Zr & 0.3-0.4  &           &          \\
$^{100}$Mo & 1.1-1.2 &         &         \\
$^{116}$Cd &  1.2-1.4 &        &         \\
$^{130}$Te  &   1.3     &  2.13 &   1.64 \\
$^{136}$Xe  &    0.6-1.0 & 1.77 & 1.36 \\
\hline
\end{tabular}}
\label{tab_nme}
\end{center}
\end{table}

Once the nuclear matrix elements are fixed (by choosing your favorite set of results),
they can be combined with the phase space factors (a complete list is available, e.g.
in the monograph\cite{BV92}) to obtain a half-life prediction for any value of the effective
mass   $\langle m_{\beta \beta} \rangle$.
It turns out that for a fixed $\langle m_{\beta \beta} \rangle$
the half-lives of different candidate nuclei do not differ very much
from each other (not more than by factors $\sim 3$ or so)
and, for example, the boundary between the degenerate and inverted hierarchy mass
regions corresponds to half-lives $\sim10^{27}$years. Thus, the next generation
of experiments should reach this region using several candidate
nuclei, making the corresponding conclusions less nuclear model dependent.

\section{Summary}

In this talk I discussed the status of 
neutrino mass determination, in particular the role of the double beta decay.
I have shown that if one makes
the minimum assumption that the light neutrinos familiar from the oscillation
experiments, which are interacting only by the left-handed weak current, are Majorana
particles, then the rate of the $0\nu\beta\beta$ decay can be related to the
absolute scale of the neutrino mass in a straightforward way.
                                                                                                      
On the other hand, it is also possible that the $0\nu\beta\beta$ decay is mediated
by the exchange of heavy particles. I explained that if  the corresponding mass
scale of such hypothetical particles is $\sim$ 1 TeV, the corresponding $0\nu$
decay rate could be comparable to the decay rate associated with the exchange
of a light neutrino. I further argued that the study of the lepton flavor violation
involving $\mu \to e$ conversion and $\mu \to e + \gamma$ decay may be used
as a ``diagnostic tool" that could help to decide which of the possible mechanisms
of the $0\nu$ decay is dominant.
                                                                                                      
Further, I have shown that the the range of the effective masses $\langle m_{\beta\beta} \rangle$
can be roughly divided into three regions of interest, each corresponding to a different
neutrino mass pattern. The region of   $\langle m_{\beta\beta} \rangle \ge$ 0.1 eV corresponds
to the degenerate mass pattern. Its exploration is well advanced, and one can rather
confidently expect that it will be explored by several $\beta\beta$ decay experiments in
the next 3-5 years. This region of neutrino masses (or most of it)
is also accessible to studies using the ordinary $\beta$ decay and/or the observational cosmology.
Thus, if the nature is kind enough to choose this mass pattern, we will have a multiple ways
of exploring it.
                                                                                                      
The region of $0.01 \le \langle m_{\beta\beta} \rangle \le 0.1$ eV
is often called the "inverted mass
hierarchy" region. In fact, both the inverted and the quasi-degenerate but normal mass orderings
are possible in this case, and experimentally indistinguishable.
Realistic plans to explore this region
using the $0\nu\beta\beta$ decay exist, but correspond to a longer time scale of about 10 years.
They require much larger, $\sim$ ton size $\beta\beta$ sources and correspondingly even
more stringent background suppression.

Intimately related to the extraction of $\langle m_{\beta\beta} \rangle$ from the decay rates is
the problem of nuclear matrix elements. At present, there is no consensus among the nuclear
theorists about their correct values, and the corresponding uncertainty. I argued that the
uncertainty is less than some suggest, and that the closeness of the Quasiparticle Random
Phase Approximation (QRPA) and Shell Model (NSM) results are encouraging. But this
is still a problem that requires further improvements.

\end{document}